\documentclass[aps,prd,twocolumn]{revtex4-1}
\usepackage{amsmath,amssymb,slashed,graphicx}
\usepackage[colorlinks=true,linkcolor=blue,citecolor=blue,urlcolor=blue]{hyperref}
\usepackage{color}

\newcommand\Ncolor{N_{\text{c}}}
\newcommand\Nflavor{N_{\text{f}}}
\newcommand\tr{\mathop{\mathrm{tr}}}
\newcommand\Tr{\mathop{\mathrm{Tr}}}
\newcommand\Det{\mathop{\mathrm{Det}}}
\newcommand\pbr[1]{\left({#1}\right)}
\newcommand\bbr[1]{\left[{#1}\right]}
\newcommand\SWZW{S_{\text{WZW}}}

\begin{document}

\title{Chiral magnetic effect in the hadronic phase}
\author{Shota~Imaki}
\email{imaki@nt.phys.s.u-tokyo.ac.jp}
\affiliation{Department of Physics, The University of Tokyo, 7-3-1 Hongo, Bunkyo-ku, Tokyo 113-0033, Japan}
\date{\today}

\begin{abstract}
	We study the chiral magnetic effect (CME) in the hadronic phase.
	The CME current involves pseudoscalar mesons to modify its functional form.
	This conclusion is independent of microscopic details.
	The strength of the CME current in the hadronic phase would decrease for two flavors.
\end{abstract}

\maketitle

\section{Introduction}
\label{sec:intro}

Among a rich variety of quantum phenomena driven by a magnetic field~\cite{Miransky:2015ava,Fukushima:2018grm},
the chiral magnetic effect (CME)~\cite{Fukushima:2008xe},
that is, the generation of an electric current along a magnetic field
in chirally imbalanced systems,
is notable for its salient characteristics.
By virtue of its anomalous origin,
the formula of the CME current is topologically protected
even in the strong coupling limit~\cite{Adler:1969gk,*Bell:1969ts,Adler:1969er},
as substantiated by holographic approaches%
~\cite{Yee:2009vw,Rebhan:2009vc,Gynther:2010ed,Gorsky:2010xu}.
As the CME persists in the long-wavelength regime,
it modifies the hydrodynamic and kinetic descriptions of chiral fluids%
~\cite{Son:2009tf,Son:2012wh,Stephanov:2012ki,Son:2012zy}.
The nondissipative nature of the CME,
which is concisely understood in terms of the time-reversal symmetry~\cite{Kharzeev:2011ds},
is also unusual and
enables one to determine the corresponding transport coefficient
in the chiral hydrodynamics~\cite{Son:2009tf}.

Heavy ion collision experiments provide experimental probes to study the CME.
(See also the realizations in
Weyl semimetals~\cite{Son:2012bg,Zyuzin:2012tv,Basar:2013iaa,Li:2014bha,Huang:2015eia,Shekhar:2015rqa}
and
lattice simulations~\cite{Buividovich:2009wi,Yamamoto:2011gk,Yamamoto:2011ks,Mueller:2016ven,Mace:2016shq}.)
This is because
ions passing near each other would generate an intense magnetic field of order
$eB \sim m_\pi^2$~\cite{Kharzeev:2007jp,Skokov:2009qp,Kharzeev:2015znc}
and the fluctuating gauge topology generates
the chirality imbalance~\cite{Kharzeev:2004ey,Kharzeev:2007tn,Kharzeev:2009fn}.
The STAR Collaboration at the Relativistic Heavy Ion Collider (RHIC)
and the ALICE Collaboration at the Large Hadron Collider
have reported the charge-dependent azimuthal correlators
which are qualitatively consistent with the charge separations caused by the CME%
~\cite{Abelev:2009ac,Abelev:2009ad,Adamczyk:2014mzf,Abelev:2012pa}.
The signals of the CME-driven collective excitation,
called a chiral magnetic wave~\cite{Kharzeev:2010gd,Burnier:2011bf,Gahramanov:2012wz,Burnier:2012ae,Stephanov:2013tga},
is also expected to be observed as charge-dependent elliptic flows~\cite{Adamczyk:2015eqo,Adam:2015vje}.
The beam energy scan program at RHIC
is continuing to examine the energy dependence of charge separations
toward the low-energy regime~\cite{Kumar:2013cqa,Aggarwal:2010cw,Adamczyk:2017iwn}.
Further understanding would be achieved by
these on-going experiments
as well as
firm quantification of the transport properties~\cite{Fukushima:2010zza,Kharzeev:2009pj,Fukushima:2017lvb,Hattori:2016cnt,Hattori:2016lqx,Fukushima:2015wck,Hattori:2017qih,Aarts:2007wj,Ding:2010ga,Ding:2016hua}
and the real-time dynamics~\cite{Fukushima:2010vw,McLerran:2013hla,Tuchin:2015oka,Mueller:2016ven,Copinger:2018ftr}
of matter under a strong magnetic field.

One complication of heavy ion collisions is that
created fireballs would undergo hadronization.
Since the fundamental degrees of freedom change
from quarks and gluons into hadrons,
one cannot directly employ
the formula of the CME derived in the chiral phase
once the system has hadronized.
However, the CME in the hadronic phase has not been studied in much detail.
(Exceptions include Ref.~\cite{Fukushima:2012fg}.)
This is one of the paramount issues,
especially when scanning the low-beam-energy regions
where quark-gluon plasma would have a short lifetime or even not be generated.
In order to examine the CME signals in heavy ion collisions with low energy,
it is indispensable to quantify the CME in the hadronic phase.

In this work we present two conclusions.
First,
the CME current in the hadronic phase is, at the functional level,
modified from the chiral phase
by involving the pseudoscalar mesons.
The current reads as Eq.~\eqref{eq:jhadron}.
Its functional form,
derived from the Wess-Zumino-Witten (WZW) action
capturing the anomalous coupling between the pseudoscalar mesons and gauge fields,
is independent of microscopic details
because the WZW action is
determined nonperturbatively by matching the anomaly.
The involvement of the pseudoscalar mesons in the CME is not quite unnatural.
Although the absence of higher-order corrections in the anomalous current has been
perturbatively proved~\cite{Adler:1969er},
has not been shown non-perturbatively.
Indeed, the anomalous triangle diagram can, in general,
involve the pseudoscalar particles
as exemplified by the $\pi^0 \to \gamma\gamma$ decay
and the Primakoff effect.
Second,
the involvement of the pseudoscalar mesons
would reduce the strength of the CME by a few percent.
The reduction is quantified by the generalized dielectric constant, Eq.~\eqref{eq:kappa},
incorporating the medium effect of the pseudoscalar mesons.
We will analytically demonstrate this effect for the two-flavor free pion gas
in equilibrium.

This paper is organized as follows.
We first review the CME in the chiral phase in Sec.~\ref{sec:chiral}.
We compute the effective action to get the renowned CME formula.
We then study the CME in the hadronic phase in Sec.~\ref{sec:hadron}.
The CME is first derived by means of a chiral effective model
to clarify the physical picture.
Afterwards, we show the model independence of the result.
It is clarified that the CME in the hadronic phase
involves the pseudoscalar mesons to modify its functional form.
We then analyze
how this modification influences the strength of the CME current
in Sec.~\ref{sec:strength}.
We carry out an analytical evaluation of the current strength
by limiting ourselves to a two-flavor free pion gas.
We find that the CME current would be reduced by a few percent.
The last section, Sec.~\ref{sec:summary}, is devoted to the summary and outlook.

\section{Chiral magnetic effect in the chiral phase}
\label{sec:chiral}

We outline the CME in the chiral phase.
The original work~\cite{Fukushima:2008xe} provides four methods of derivation,
among which we adopt a derivative expansion of the effective action.

We consider a system in the chiral phase
with an external magnetic field and
imbalance between right-handed and left-handed fermions.
One way to derive the current in this system is
through the effective action
\begin{align}
	S_{\text{eff}}
	= - i \log \Det (i \slashed{D} - m) \,.
	\label{eq:logdetchiral}
\end{align}
The covariant derivative
$i D_\mu = i\partial_\mu - e Q A_\mu - \gamma_5 a_\mu$
incorporates
the electromagnetic field $A_\mu$ associated with the charge matrix $Q$ and
the axial gauge field $a_\mu = (\mu_5, \boldsymbol{0})$ encoding the axial chemical potential.
The electromagnetic field strength is denoted by $F_{\mu\nu}$.
The determinant is over the coordinate space
as well as the flavor, color, and Dirac indices.
The quark masses are set to the same value $m$ for brevity.
The derivative expansion of the effective action \eqref{eq:logdetchiral}
is performed as in Ref.~\cite{DHoker:1985ycn}.
To determine the effective action without the renormalization scheme dependence,
it is necessary to impose the physical requirement that
the current should generate the canonical anomalous divergence
\begin{align}
	\partial_\mu j^\mu
	= \frac{e^2 \Ncolor}{16 \pi^2}
	\left(
		F^{\mathrm{R} \mu\nu} \tilde F^{\mathrm{R}}_{\mu\nu}
		- F^{\mathrm{L} \mu\nu} \tilde F^{\mathrm{L}}_{\mu\nu}
	\right) \,,
\end{align}
with $\Ncolor$ being the number of colors.
This requirement identifies the effective action as
\begin{align}
	S_{\text{eff}}
	= \frac{e^2 \Ncolor}{4 \pi^2} \int d^4x \;
	a_\mu A_\nu \tilde F^{\mu\nu} \tr(Q^2)
	\label{eq:effchiral}
\end{align}
in the leading order of the derivative expansion.
We omitted the terms that are irrelevant to the CME.
A functional derivative of Eq.~\eqref{eq:effchiral} with respective to $\boldsymbol{A}$
yields the CME current
\begin{align}
	\boldsymbol{j}
	= \frac{e^2 \Ncolor}{2\pi^2} \mu_5 \boldsymbol{B} \tr(Q^2) \,.
	\label{eq:jchiral}
\end{align}
We note that this derivation as such assumes a static and homogeneous magnetic field.

\section{Chiral magnetic effect in the hadronic phase}
\label{sec:hadron}

We have observed that the CME in the chiral phase
is given by Eq.~\eqref{eq:jchiral}.
In this section we derive the CME in the hadronic phase.
The derivation is performed in two ways.
First, we adopt a chiral effective model to get the effective action.
This method clarifies the physical picture of the result.
Second, we argue that the WZW action~\cite{Wess:1971yu,*Witten:1983tw}
gives the identical effective action.
This method verifies the model independence of the result.
The obtained effective action will yield the CME current.

\begin{figure}[t]
	\begin{center}
		\includegraphics[width=.24\linewidth]{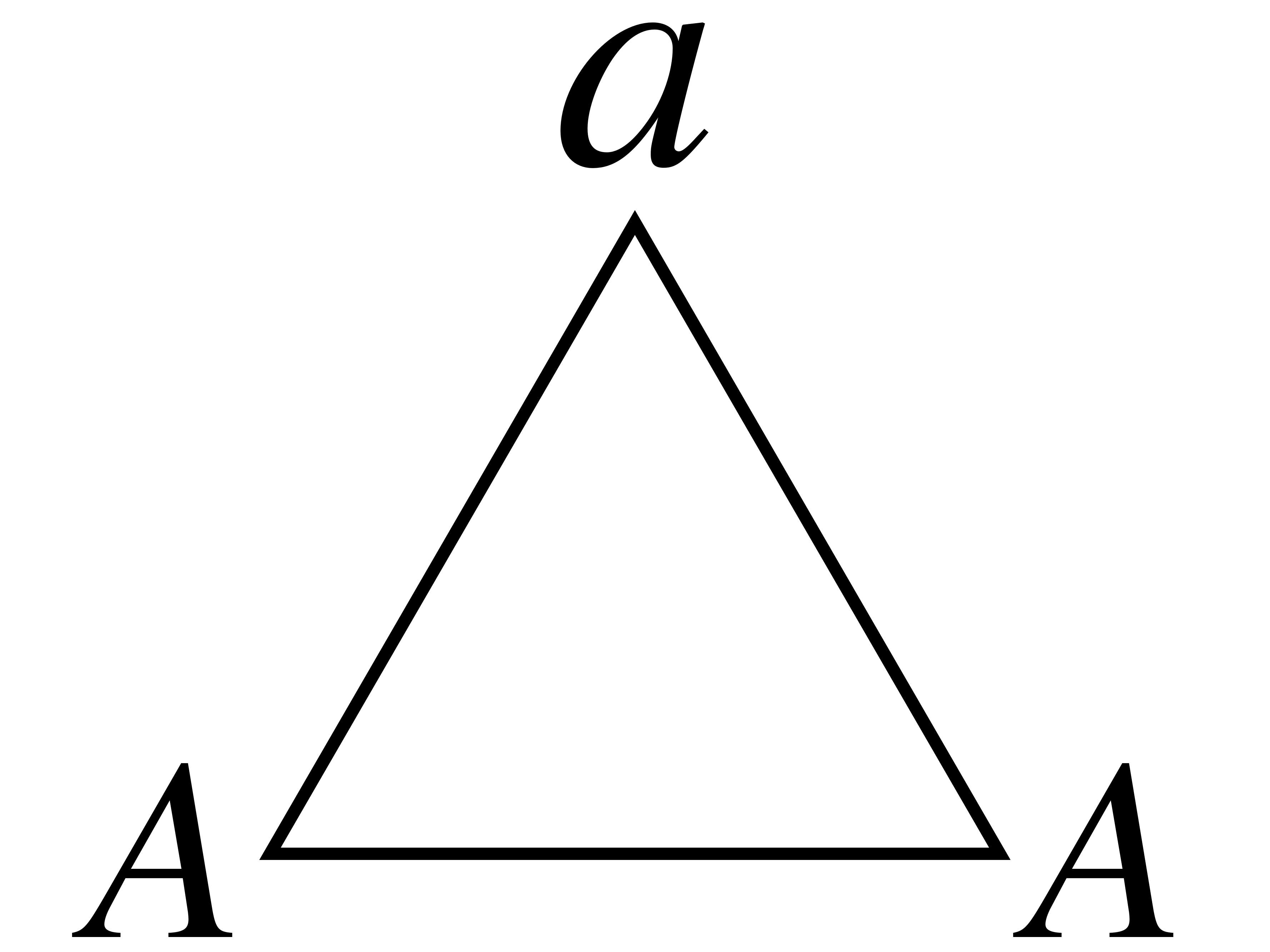}
		\includegraphics[width=.24\linewidth]{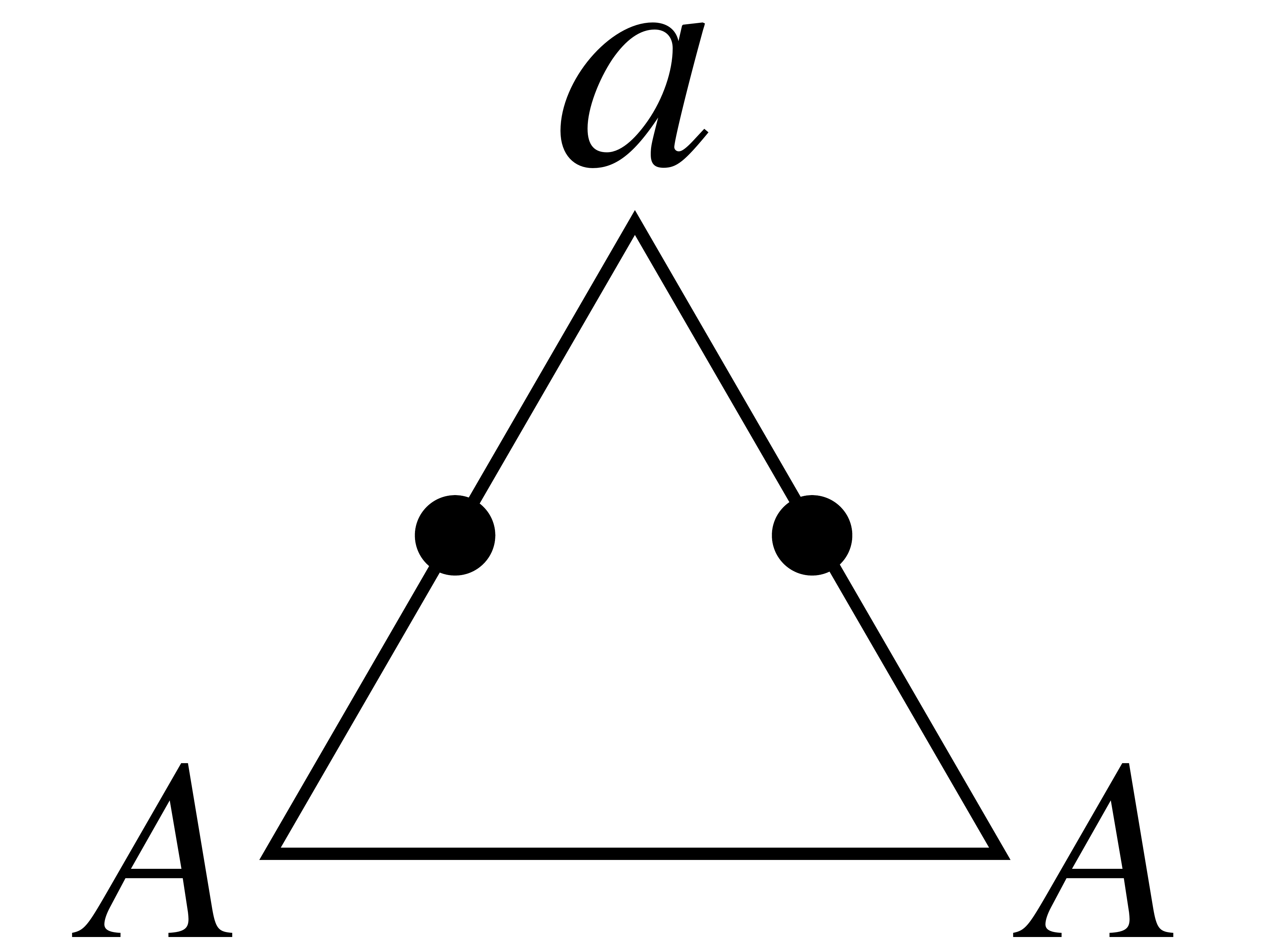}
		\includegraphics[width=.24\linewidth]{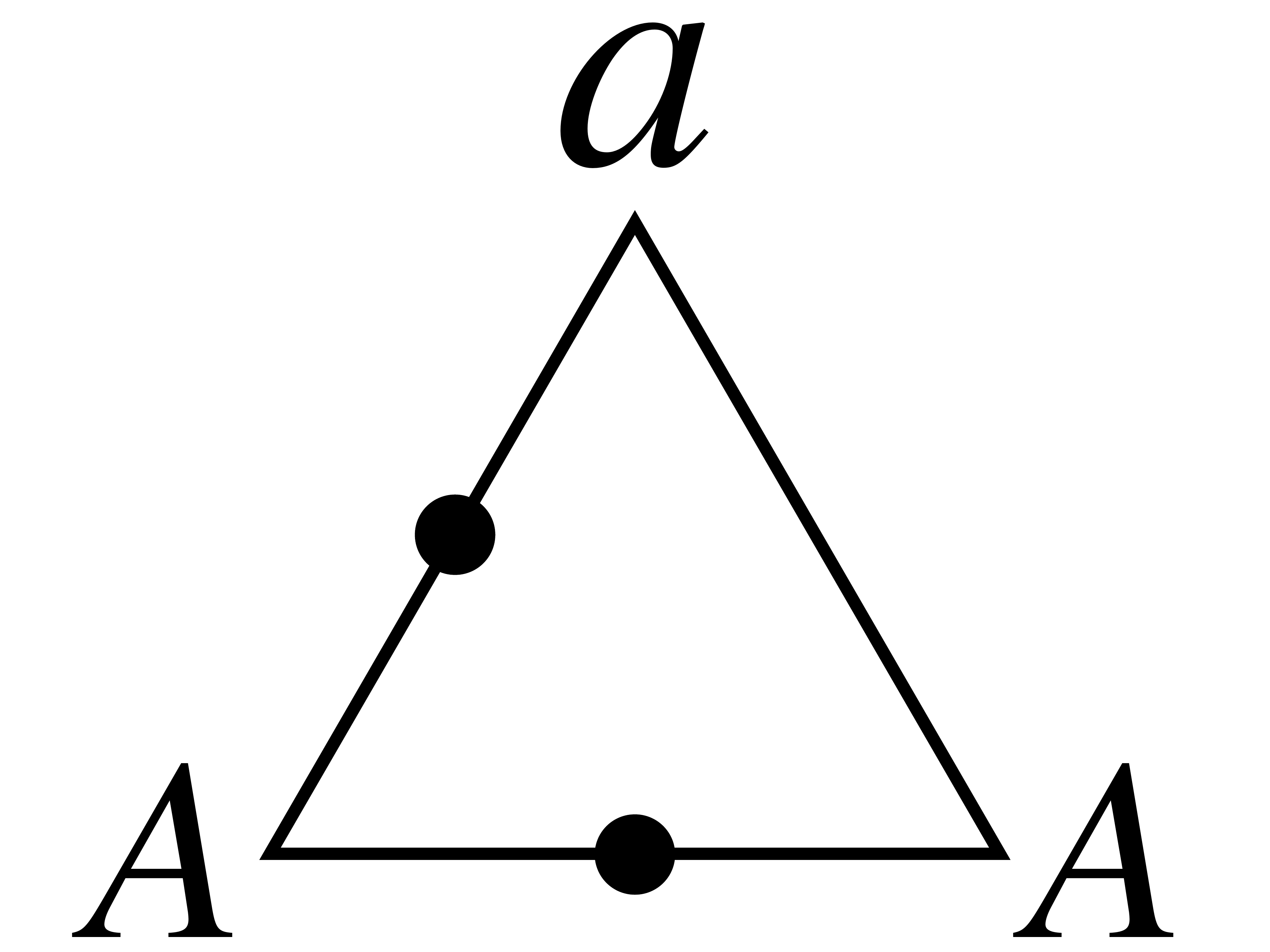}
		\includegraphics[width=.24\linewidth]{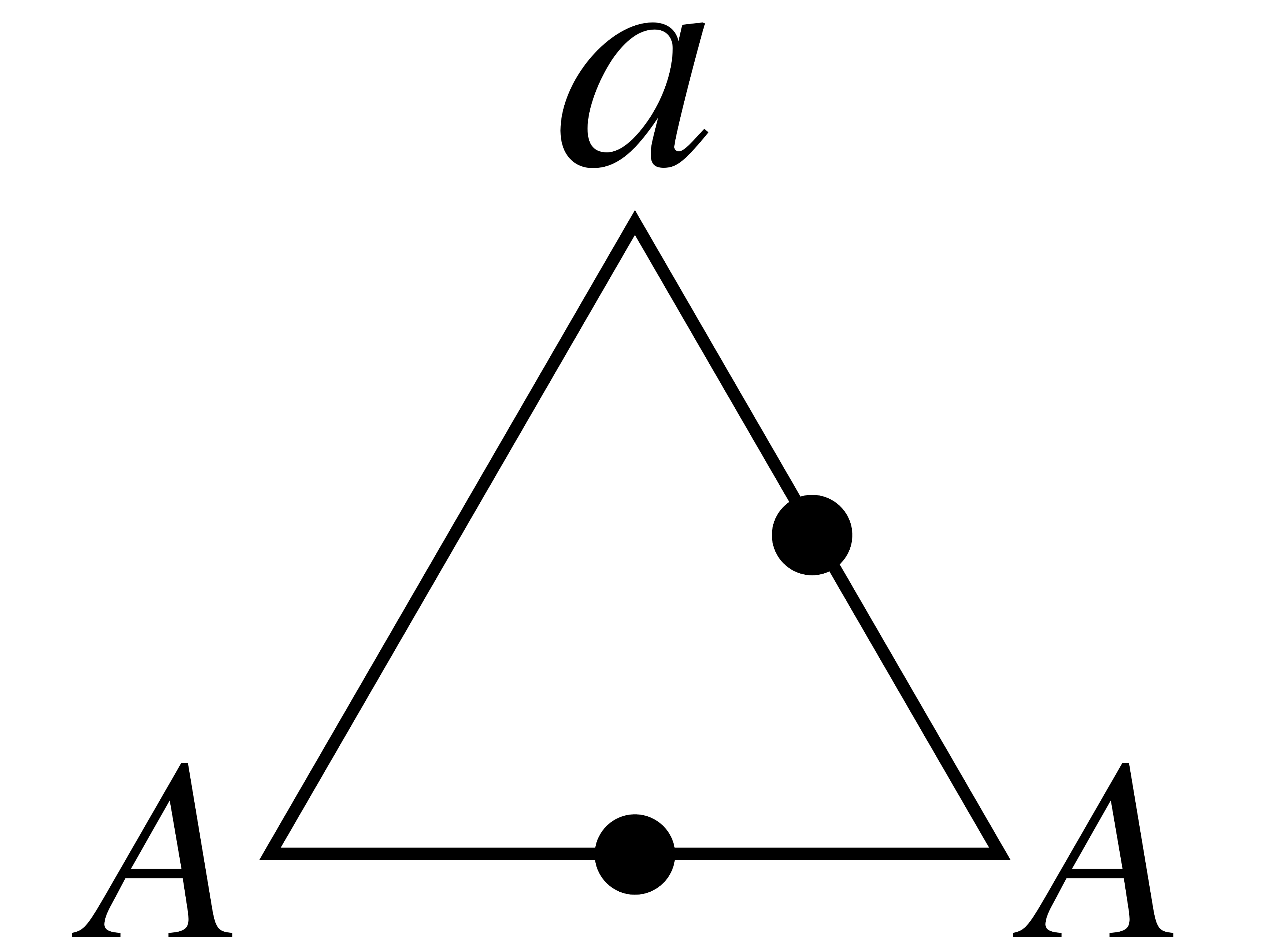}
		\caption{
			The triangle diagrams for the CME in the hadronic phase.
			The black dots represent the insertion of $M$ in Eq.~\eqref{eq:trloghadron}.
			\label{fig:triangle}
		}
	\end{center}
\end{figure}

Now, we consider a system in the hadronic phase
with an external magnetic field and the chiral chemical potential.
The chiral chemical potential can be substituted by the oscillating vacuum angle
$\theta = 2 N_{\text{f}} \mu_5 t$
or temporal fluctuation of the pseudoscalar mesons.
Since the realizations of $\mu_5$ bring about the time dependence of the system,
the following computations of the current should be justified away from equilibrium
by means of the Kubo formalism or real-time formalism.
We first adopt the chiral effective model
called the nonlinear quark-meson model~\cite{Kamikado:2012bt,Schaefer:2004en}.
This model properly realizes the chiral symmetry pattern in the hadronic phase,
in terms of the pseudoscalar meson multiplet
$\Sigma \equiv \exp (i \pi^A \lambda^A / f_\pi)$
with $A$ being the adjoint flavor indices,
and thus reproduces consistent results in the low-energy limit.
The pseudoscalar mesons are the background fields for the moment
while we will treat them as the dynamical fields in Sec.~\ref{sec:strength}.
The Lagrangian of the quark sector reads
\begin{align}
	\mathcal{L}
	= \bar q (i \slashed{D} - g M) q \,,
\end{align}
where $M = P_{\mathrm{R}} \Sigma + P_{\mathrm{L}}\Sigma^\dagger$
with $P_{\mathrm{R}}$ and $P_{\mathrm{L}}$ being chiral projectors
while $g \neq 0$ denotes the coupling constant.
The quarks $q$ are constituent ones.
The covariant derivative incorporates
the magnetic field and the axial chemical potential
as in Eq.~\eqref{eq:logdetchiral}.
To derive the electric current, we again seek the effective action
\begin{align}
	S_{\text{eff}}
	= - i \log \Det (i\slashed{D} - g M) \,.
	\label{eq:logdethadron}
\end{align}
Note that this effective action formally reduces to Eq.~\eqref{eq:logdetchiral}
if the pseudoscalar mesons are absent, namely, $\Sigma = 1$.
The perturbative expansion of the effective action yields
\begin{align}
	\begin{split}
	&S_{\text{eff}}
	= - i e^2 \\
	& \cdot \Tr \pbr{
		\gamma_5 \slashed{a}
		\frac{i \slashed\partial + g M^\dagger}{- \partial^2 - g^2}
		Q \slashed{A}
		\frac{i \slashed\partial + g M^\dagger}{- \partial^2 - g^2}
		Q \slashed{A}
		\frac{i \slashed\partial + g M^\dagger}{- \partial^2 - g^2}
	} \,,
	\label{eq:trloghadron}
	\end{split}
\end{align}
among which the nonvanishing contributions are depicted by the triangle diagrams in Fig.~\ref{fig:triangle}.
We omitted the terms that are irrelevant to the CME.
In contrast to the chiral phase,
the effective action involves the pseudoscalar mesons
through $M$ in Eq.~\eqref{eq:trloghadron}.
Further computation can be performed
by imposing the physical requirement that the effective action
should reduce to Eq.~\eqref{eq:effchiral} for $\Sigma = 1$.
The result,
in the leading order of the derivative expansion,
reads
\begin{align}
	S_{\text{eff}}
	&= \frac{e^2 \Ncolor}{12 \pi^2} \int d^4x \;
	a_\mu A_\nu \tilde F^{\mu\nu}
	\tr (2 Q^2 + Q \Sigma Q \Sigma^\dagger) \,,
	\label{eq:effhadron}
\end{align}
which is independent of $g$.
The derivation is given in Appendix~\ref{app:eff}.

The effective action \eqref{eq:effhadron}
is also obtained from the WZW action
derived in Appendix~\ref{app:WZW}.
This argument qualifies the effective action \eqref{eq:effhadron} to be
independent of microscopic details in the low-energy regime.

The effective action \eqref{eq:effhadron} yields the CME current
\begin{align}
	\boldsymbol{j}
	= \frac{e^2 \Ncolor}{2 \pi^2} \mu_5 \boldsymbol{B}
	\tr \pbr{Q^2 + \frac16 [Q, \Sigma] [Q, \Sigma^\dagger]} \,.
	\label{eq:jhadron}
\end{align}
This is our main result.
One can clearly see that the CME current involves the pseudoscalar mesons
to modify its form from that in the chiral phase \eqref{eq:jchiral}.
The topological nature of the WZW action implies that
the functional form of the current \eqref{eq:jhadron}
does not receive higher-loop corrections.
Note that this formula is for the pseudoscalar mesons being background fields.
After taking these vacuum or thermal expectation values
upon a proper redefinition of physical quantities,
Eq.~\eqref{eq:jhadron} reduces to a similar form as the chiral phase.
We will substantiate this argument in the next section.

Before leaving this section,
we make a remark on the disagreement with
the work by Fukushima and Mameda~\cite{Fukushima:2012fg}.
Their work demonstrates that
the CME current in the hadronic phase maintains
the same functional form as the chiral phase.
This disagreement stems from the WZW action that
they cite from Ref.~\cite{Kaiser:2000gs,*Kaiser:2000ck}.
The WZW action in Ref.~\cite{Kaiser:2000gs,*Kaiser:2000ck}
does not have the term with the pseudoscalar mesons in Eq.~\eqref{eq:effhadron},
being contrary to the one derived in Appendix~\ref{app:WZW}
and the multiple preceding works~\cite{Chou:1983qy,*Kawai:1984mx,*Kaymakcalan:1983qq,*Manes:1984gk}.
Although the WZW action in Ref.~\cite{Chou:1983qy,*Kawai:1984mx,*Kaymakcalan:1983qq,*Manes:1984gk}
as such
yields a different action from Eq.~\eqref{eq:effhadron}
by a term proportional to
$a_\mu A_\nu \tilde F^{\mu\nu} \tr(Q^2)$,
this term stems from the difference of counterterms.
One can tame this renormalization scheme dependence
by the requirement that
the current should generate the one in the chiral phase.

\section{Strength of the CME current}
\label{sec:strength}

We have shown that the CME current in the hadronic phase \eqref{eq:jhadron}
involves the pseudoscalar mesons.
Now our interest is in how much this effect modifies the strength of the current.
We examine this issue, for simplicity,
by limiting ourselves to a two-flavor free pion gas at finite temperature in equilibrium.
The extension to the nonequilibrium case requires appropriate treatment of the chirality imbalance
as in Refs.~\cite{Mueller:2016ven,Mace:2016shq}.
This limitation enables an analytical evaluation of the current strength.
We will see that the current strength is decreased by a few percent.

We hereby treat the pions as the dynamical fields.
In this case, the flow of the charged pions also carries the electric current,
but we ignore this nonanomalous contribution.
The pions are assumed to be free with their mass $m_\pi$ and decay constant $f_\pi$
being substituted by the thermal effective values
given in Ref.~\cite{Petropoulos:1998gt}.
The magnetic-field dependences of $m_\pi$ and $f_\pi$ are ignored.
The vacuum or thermal expectation value of the CME current \eqref{eq:jhadron} reads
\begin{align}
	\langle \boldsymbol{j} \rangle
	= \kappa \cdot \frac{e^2 \Ncolor}{2 \pi^2} \mu_5 \boldsymbol{B} \tr (Q^2) \,,
\end{align}
where we defined
\begin{align}
	\kappa
	\equiv \left\langle \tr \pbr{Q^2 + \frac16 [Q, \Sigma] [Q, \Sigma^\dagger]} \middle\rangle
	\right/ \tr(Q^2) \,.
	\label{eq:kappa}
\end{align}
The bracket $\langle\cdots\rangle$ denotes the vacuum or thermal expectation value.
The coefficient $\kappa$ is interpreted as the generalized dielectric constant
incorporating the medium effect of the pseudoscalar mesons~\cite{Fukushima:2010zza,Kharzeev:2009pj}.
In other words,
the magnetic field would be substituted by $\boldsymbol{H} = \kappa \boldsymbol{B}$
in medium, as suggested in Ref.~\cite{Fukushima:2010zza}.
With this redefinition the anomaly relation \eqref{eq:jchiral} still holds.

As shown in Appendix.~\ref{app:tr},
the dielectric constant \eqref{eq:kappa}
is evaluated in terms of the thermal Green function
at the coincidental point~\cite{laine2016basics}
\begin{align}
	\begin{split}
	G
	& \equiv f_\pi^{-2} \langle \pi^A(x) \pi^A(x) \rangle \\
	& = G_0 - \frac{m_\pi^2}{16 \pi^2 f_\pi^2}
	+ \frac{m_\pi T}{2 \pi^2 f_\pi^2} \sum_{n = 1}^\infty
	\frac{1}{n} K_1 \pbr{\frac{m_\pi n}{T}} \,,
	\end{split}
	\label{eq:G}
\end{align}
where $K_1(z)$ denotes the modified Bessel function of the second kind.
The sum over $A = 1, 2, 3$ is not taken here.
The constant $G_0$ is the counterterm to be determined shortly.
The result reads
\begin{align}
	\kappa
	= \frac15 (12 + 3 e^{- 2 G} + 9 e^{- G} - 18 e^{- \frac12 G}) \,.
	\label{eq:kappaofG}
\end{align}
In order to determine $G_0$,
we impose the requirement that the dielectric constant \eqref{eq:kappaofG}
should be unity at the temperature of chiral symmetry restoration, $T = 180\;\text{MeV}$.
This requirement gives $G_0 = 1.48\;\text{MeV}$.

\begin{figure}[t]
	\includegraphics[width=1.0\linewidth]{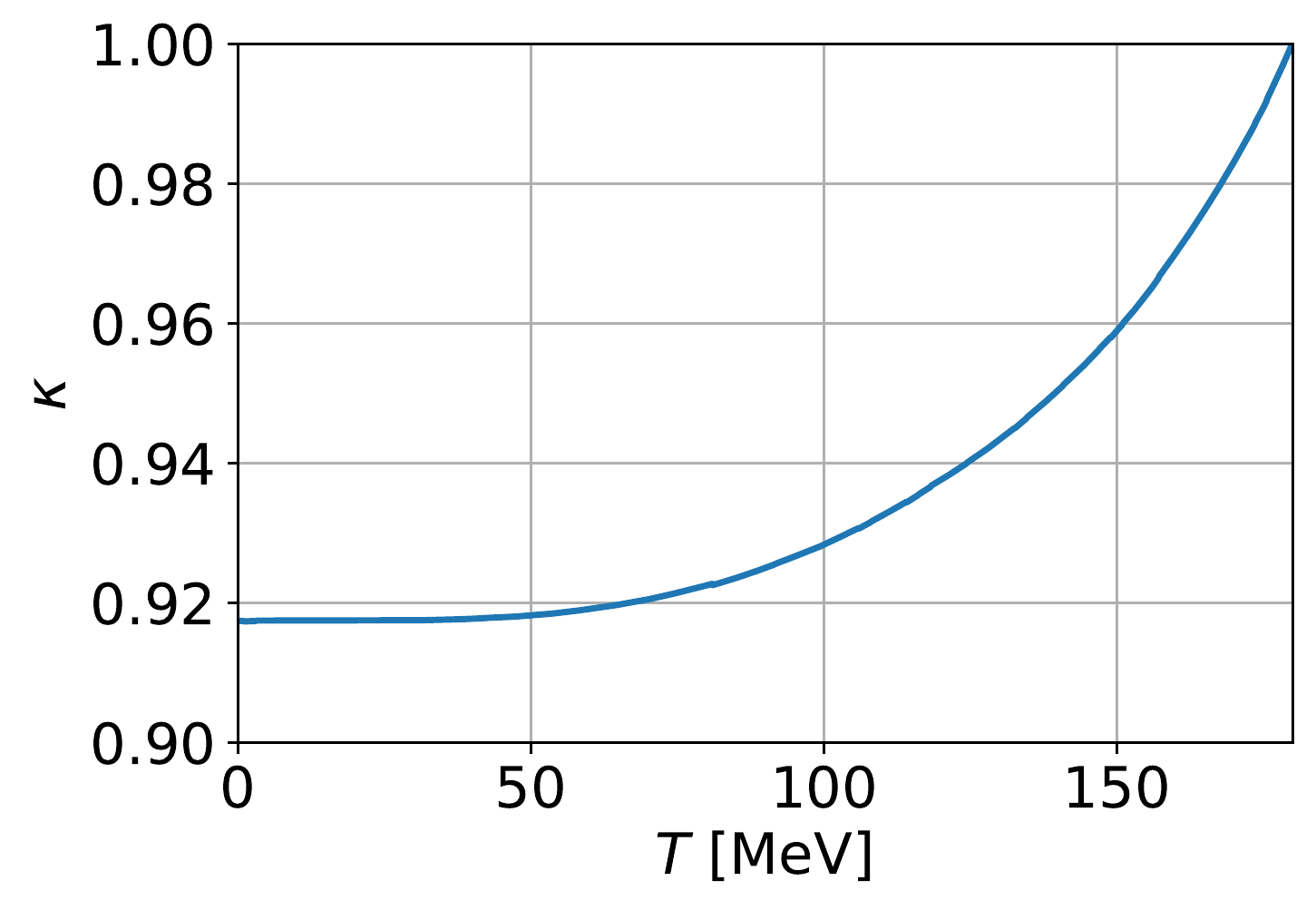}
	\caption{
		\label{fig:kappa}
		Temperature dependence of the dielectric constant.
	}
	\label{fig:f}
\end{figure}

The temperature dependence of the dielectric constant $\kappa$
is shown in Fig~\ref{fig:kappa}.
The strength of the CME current is reduced from that in the chiral phase.
It is interesting to note that
the beam energy scan programs in ALICE and STAR
have reported reduced charge separations for low beam energies~\cite{Abelev:2012pa,Adamczyk:2014mzf},
for which quark-gluon plasma would have a short lifetime until the system hadronizes.

\section{Summary and outlook}
\label{sec:summary}

We studied the CME in the hadronic phase
by means of the chiral effective model as well as the WZW action.
The CME current is given by Eq.~\eqref{eq:jhadron}
and involves the pseudoscalar mesons.
This result is independent of microscopic details.
The involvement of the pseudoscalar mesons
can either increase or decrease the strength of the CME current.
In particular, the analysis of the two-flavor case implies that
the CME signal would be reduced when the collision energy is so low
that the system quickly hadronizes.
This result qualitatively agrees with the observations of
the beam energy scan programs.

The large multipion correlations,
which were recently reported by the ALICE Collaboration~\cite{Adam:2015pbc,Begun:2016cva},
might considerably influence the behavior of the CME signals.
A recent theoretical study implied that
these large correlations could be the manifestation of
the Bose-Einstein condensation of the charged pions~\cite{Liu:2017spl}.
(See also Refs.~\cite{Brauner:2016pko,Huang:2017pqe} for related works.)
Although we have limited ourselves to a free pion gas in Sec.~\ref{sec:strength},
further analysis including the multipion correlations
is necessary to inspect these interesting phenomena.

Other chiral transport phenomena,
most of which have been examined in the chiral phase,
could be enriched by hadronic environments.
For instance,
it is clear from Eq.~\eqref{eq:effhadron} that the chiral separation effect (CSE)
also involves the pseudoscalar mesons.
Accordingly, the chiral magnetic wave,
which is derived by combining the CME and the CSE,
would change its behavior in the hadronic phase.
Besides, one could examine the chiral vortical effect and
the chiral torsional effect~\cite{Khaidukov:2018oat,Imaki:2019ite}
in the hadronic phase
by incorporating the pseudoscalar mesons.
These intriguing transport phenomena deserve further investigations.

\begin{acknowledgments}
	The author is grateful to Kenji~Fukushima, Misha~Stephanov and Ho-Ung~Yee for discussions.
	The author is supported by Grant-in-Aid for JSPS Fellows Grant Number 19J22323.
\end{acknowledgments}

\appendix

\section{Derivation of Eq.~\eqref{eq:effhadron}}
\label{app:eff}

We derive the effective action \eqref{eq:effhadron}.
We neglect the momenta of the pseudoscalar mesons in what follows.
The perturbative expansion of the effective action \eqref{eq:logdethadron} yields
\begin{align*}
	S_{\text{eff}}
	&= - i \Ncolor \Tr \log (i \slashed\partial - g M) \\
	&\quad
	- i \Ncolor \sum_{n = 1}^\infty \frac{1}{n}
	\Tr \bbr{
		\pbr{\frac{i \slashed\partial + g M^\dagger}{- \partial^2 - g^2}
		(e Q \slashed A + \gamma_5 \slashed a)}^n
	} \,.
\end{align*}
The trace is over the coordinate space as well as the flavor and Dirac indices.
As we are interested in the CME,
we focus on the term that is linear in $a_\mu$ and quadratic in $A_\mu$,
which is from $n = 3$.
This term reads, in the momentum space,
\begin{align}
	S_{\text{eff}}
	= - i \Ncolor \int \frac{d^4 p}{(2\pi)^4} \;
	a_\mu A_\nu(p) A_\rho(-p) K^{\mu\nu\rho}(p) \,.
	\label{eq:effK}
\end{align}
The kernel $K^{\mu\nu\rho}(p)$ is given by
the triangle diagrams in Fig.~\ref{fig:triangle}, or, the integral
\begin{align*}
	& K^{\mu\nu\rho}(p)
	= e^2 \int \frac{d^4 k}{(2\pi)^4} \; \\
	& \cdot
	\tr \bigg(
		\gamma_5 \gamma^\mu
		\frac{\slashed{k} + g M^\dagger}{k^2 - g^2}
		Q \gamma^\nu
		\frac{\slashed{k} + \slashed{p} + g M^\dagger}{(k + p)^2-g^2}
		Q \gamma^\rho
		\frac{\slashed{k} + g M^\dagger}{k^2 - g^2}
	\bigg) \,.
\end{align*}
By virtue of the trace identity of gamma matrices,
only the terms involving even numbers of $M$ are nonvanishing.
The term with two $M$'s is given by a convergent integral and reads
\begin{align*}
	K^{\mu\nu\rho}_1(p)
	&= - \frac{e^2}{12 \pi^2} \epsilon^{\mu\nu\rho\sigma} p_\sigma
	\tr (Q^2 + Q \Sigma Q \Sigma^\dagger) \,,
\end{align*}
in the leading order of the derivative expansion.
On the other hand, the term without an $M$
is divergent and thus depends on regularization schemes.
We now impose a requirement that
the CME current with $\Sigma = 1$ must reproduce that in the chiral phase
\eqref{eq:jchiral}.
This requirement identifies the coefficient with the form
\begin{align*}
	K^{\mu\nu\rho}(p)
	= - \frac{e^2}{12 \pi^2} \epsilon^{\mu\nu\rho\sigma} p_\sigma
	\tr (2 Q^2 + Q \Sigma Q \Sigma^\dagger) \,.
\end{align*}
By plugging this into Eq.~\eqref{eq:effK},
we reach the effective action
which is given by Eq.~\eqref{eq:effhadron} in the coordinate space.

\section{Derivation of the WZW action}
\label{app:WZW}

We derive the WZW action which leads to Eq.~\eqref{eq:effhadron}.
We adopt the notation of differential forms.
The WZW action is used to reproduce the anomaly,
\begin{align}
	i \Delta (\alpha, \beta) \SWZW [v, a, \Sigma]
	= - \int d^4 x \; \tr (\beta \mathcal{A} [v, a]) \,,
	\label{eq:bardeen}
\end{align}
with $\Delta$ being the generator of an infinitesimal chiral transformation.
The anomaly reads~\cite{Bardeen:1969md}
\begin{align*}
	& \mathcal{A} [v, a] \\
	& = \frac{\Ncolor}{4 \pi^2}
	\bbr{
		f_v^2 + \frac{1}{3} (D_v a)^2
		+ \frac{i}{3} (f_v a^2 + 4 a f_v a + a^2 f_v) + \frac{1}{4} a^4
	} \,,
\end{align*}
where $v$ and $a$ are the vector and axial gauge fields
of the flavor $\mathrm{U}(\Nflavor)$ group, respectively.
We denote $f_v \equiv dv - i v^2$ and $D_v a \equiv da - i v a - i a v$.
Equation~\eqref{eq:bardeen} implies
\begin{align*}
	& [i \Delta (\alpha, \beta)]^n \SWZW [v, a, \Sigma] \\
	& = - \int d^4 x \;
	[i \Delta (\alpha, \beta)]^{n - 1}
	\tr (\beta \mathcal{A} [v, a]) \,,
\end{align*}
which gives
\begin{align*}
	& e^{i \Delta (\alpha, \beta)} \SWZW [v, a, \Sigma] \\
	& = \SWZW [v, a, \Sigma]- \int d^4 x \;
	\frac{e^{i \Delta (\alpha, \beta)} - 1}{i \Delta (\alpha, \beta)}
	\tr (\beta \mathcal{A} [v, a]) \\
	& = \SWZW [v, a, \Sigma]- \int_0^1 ds \int d^4 x \;
	e^{i \Delta (s \alpha, s \beta)}
	\tr (\beta \mathcal{A} [v, a]) \,.
\end{align*}
For $\alpha = \alpha_\Sigma$ and $\beta = \beta_\Sigma$
satisfying the condition
\begin{align}
	e^{i (\alpha_\Sigma + \beta_\Sigma)} \Sigma e^{- i (\alpha_\Sigma - \beta_\Sigma)}
	= 1
	\label{eq:alphasigma}
	\,,
\end{align}
the relation above gives rise to
\begin{align}
	\begin{split}
	& \SWZW [v, a, \Sigma] \\
	& = \SWZW [v, a, 1]
	+ \int_0^1 ds \int d^4 x \;
	\tr (\beta \mathcal{A} [v_s, a_s]) \,.
	\end{split}
	\label{eq:wzwsolution}
\end{align}
We defined the fields $v_s$ and $a_s$ by the equations
\begin{align}
	(v_s \pm a_s)
	& \equiv e^{i s (\alpha_\Sigma \pm \beta_\Sigma)}
	(v \pm a + i d)
	e^{- i s (\alpha_\Sigma \pm \beta_\Sigma)} \,.
	\label{eq:vsas}
\end{align}
Although $\alpha_\Sigma$ and $\beta_\Sigma$
satisfying the condition \eqref{eq:alphasigma} are not unique,
any of these yield the unique WZW action \eqref{eq:wzwsolution}
by virtue of the Wess-Zumino consistency condition~\cite{Wess:1971yu}.
With this understanding, we choose the pair of parameters as
\begin{align*}
	\alpha_\Sigma = - \beta_\Sigma \,, \quad
	e^{- 2 i \beta_\Sigma} = \Sigma \,.
\end{align*}
The choice of $\SWZW[v, a, 1]$ in Eq.~\eqref{eq:wzwsolution}
corresponds to the choice of counterterms and will be made later.

We will now compute the WZW action \eqref{eq:wzwsolution}.
We split the axial gauge field into the $\mathrm{U}(1)$ component
and the remnant,
\begin{align*}
	a = \hat{a} + \frac{1}{\Nflavor} \tr(a) \,.
\end{align*}
Accordingly the anomaly is given by
\begin{align}
	\mathcal{A} [v, a]
	= \mathcal{A} [v, \hat{a}] + \tilde{\mathcal{A}} [v, a] \,,
	\label{eq:Asplit}
\end{align}
where
\begin{align*}
	\tilde{\mathcal A}[v, a]
	= \frac{\Ncolor}{6 \Nflavor \pi^2}
	\bbr{
		(D_v \hat{a}) d + 2 i (\hat a f_v - f_v \hat a)
	} \tr(a) \,.
\end{align*}
As we are interested in the CME,
we hereby limit the axial gauge field $a$ to a $\mathrm{U}(1)$ field
and focus on the terms in the WZW action that contain $a$.
This limitation simplifies the definition \eqref{eq:vsas} to
\begin{align*}
	v_s
	& = \frac12 (v + \Sigma_s v \Sigma_s^\dagger + i L_s) \,, \\
	a_s
	& = \frac12 (v - \Sigma_s v \Sigma_s^\dagger - i L_s) + a \,,
\end{align*}
with the notations
$\Sigma_s = e^{- 2 i s \beta_\Sigma}$,
$L_s \equiv \Sigma_s d \Sigma_s^\dagger$ and
$R_s \equiv d \Sigma_s^\dagger \Sigma_s$.
One may notice that $a$ appears in the WZW action
only through $\tilde{\mathcal{A}} [v_s, a_s]$ in Eq.~\eqref{eq:Asplit}.
Using the relation $\partial_s \Sigma_s = - 2 i \beta \Sigma_s$,
we find
\begin{align*}
	& \tr(\beta \tilde{\mathcal{A}} [v_s, a_s]) \\
	& = \frac{\Ncolor}{24 \pi^2} \partial_s
	\tr \Big[
		(\Sigma_s v \Sigma_s^\dagger + \Sigma_s^\dagger v \Sigma_s) (dv - 2 i v^2) \\
		& + v \Sigma_s v \Sigma_s^\dagger L_s + v \Sigma_s^\dagger v \Sigma_s R_s
		+ i f_v (L_s + R_s) - i v L_s L_s
	\Big] a \,.
\end{align*}
Thus one can readily carry out the integration with respect to $s$
in the formula Eq.~\eqref{eq:wzwsolution}.
The result reads
\begin{align*}
	& \SWZW [v, a, \Sigma]
	= \SWZW [v, a, 1] \\
	& + \frac{\Ncolor}{24 \pi^2} \int d^4x \;
	\tr \Big[
		(\Sigma v \Sigma^\dagger + \Sigma^\dagger v \Sigma - 2 v) (dv - 2 i v^2) \\
		& + v \Sigma v \Sigma^\dagger L + v \Sigma^\dagger v \Sigma R
		+ i f_v (L + R)
		- i v L L
	\Big] a \,.
\end{align*}
We omitted the terms without $a$.
If the vector field is the electromagnetic field, $v = Q A$,
this result becomes as simple as
\begin{align*}
	& \SWZW [Q A, a, \Sigma]
	= \SWZW [Q A, a, 1] \\
	& + \frac{\Ncolor}{24 \pi^2} \int d^4 x \;
	\Big\{
		2 A F a \tr(Q \Sigma Q \Sigma^\dagger - Q^2) \\
		& + i F a \tr[Q (L + R)] - i A a \tr(Q L L)
	\Big\} \,,
\end{align*}
with $F \equiv dA$.
The term $\SWZW [Q A, a, 1]$ is the effective action for $\Sigma = 1$
and should be Eq.~\eqref{eq:effchiral}.
This requirement determines the WZW action which gives Eq.~\eqref{eq:effhadron}.

\section{Derivation of Eq.~\eqref{eq:kappaofG}}
\label{app:tr}

We consider a two-flavor free pion gas
to derive the dielectric constant \eqref{eq:kappaofG}.
We represent the pions by $\Sigma = \exp (i \Pi^A \tau^A)$,
i.e. $\Pi^A \equiv \pi^A / f_\pi$,
and the charge matrix by $Q = q^0 + q^A \tau^A$.
The Pauli matrices are denoted by $\tau^A$ ($A = 1,\,2,\,3$).
The quantity of our interest reads
\begin{align*}
	\tr (2 Q^2 + Q \Sigma Q \Sigma^\dagger)
	= 6 q^0 q^0 + q^X q^Y (4 \delta^{XY} + T^{XY}) \,,
\end{align*}
where $T^{XY} \equiv \langle \tr (\tau^X \Sigma \tau^Y \Sigma^\dagger) \rangle$.
This trace is given by
\begin{align}
	\begin{split}
	T^{XY}
	&= \sum_{n = 0}^\infty \sum_{m = 0}^\infty
	\frac{i^n (-i)^m}{n! \, m!} \\
	& \quad \cdot
	\langle \Pi^{A_1} \cdots \Pi^{A_n} \Pi^{B_1} \cdots \Pi^{B_m} \rangle \\
	& \quad \cdot
	\tr(\tau^X \tau^{A_1} \cdots \tau^{A_n} \tau^Y \tau^{B_1} \cdots \tau^{B_m}) \,.
	\end{split}
	\label{eq:TXY}
\end{align}
The bosonic nature of pions and the anticommutative nature of Pauli matrices
imply that the indices $A_1, \dots, A_n$ and $B_1, \dots, B_m$ in Eq.~\eqref{eq:TXY}
respectively must take the same values.
This observation simplifies this sum as
\begin{align}
	\begin{split}
	T^{XY}
	&= \sum_{n = 0}^\infty \sum_{m = 0}^\infty
	\frac{i^n (-i)^m}{n! \, m!} \\
	& \quad \cdot
	\langle (\Pi^A)^n (\Pi^B)^m \rangle
	\tr \bbr{\tau^X (\tau^A)^n \tau^Y (\tau^B)^m} \,.
	\label{eq:TXYsimplified}
	\end{split}
\end{align}
We further observe that this trace
is nonvanishing only when $n$ and $m$ are both odd or both even.
For odd $n$ and $m$,
the vacuum expectation value in Eq.~\eqref{eq:TXYsimplified}
is nonvanishing only when the indices $A$ and $B$ are equal.
Then Wick's theorem gives
\begin{align*}
	\left. T^{XY} \right|_{\text{odd} \, n, m}
	&= \sum_{n' = 0}^\infty \sum_{m' = 0}^\infty
	\frac{i^{2n' + 1} (-i)^{2 m' + 1}}{(2 n' + 1)! \, (2 m' + 1)!} \\
	& \quad \cdot
	\langle (\Pi^A)^{2 n' + 2 m' + 2} \rangle
	\tr (\tau^X \tau^A \tau^Y \tau^A) \\
	&= \delta^{XY} (e^{- 2 G} - 1) \,.
\end{align*}
The sum with even $n$ and $m$ can also be readily evaluated.
After all, we obtain
\begin{align*}
	T^{XY}
	= \delta^{XY} (10 + 4 e^{- 2 G} + 12 e^{- G} - 24 e^{- \frac12 G} ) \,.
\end{align*}
Thus, for the charge matrix of the $u$ and $d$ quarks,
$Q = \mathrm{diag} (\frac23, -\frac13)$,
\begin{align*}
	\tr (2 Q^2 + Q \Sigma Q \Sigma^\dagger)
    = \frac{1}{3} (11 + 3 e^{- 2 G} + 9 e^{- G} - 18 e^{- \frac12 G}) \,.
\end{align*}
It gives rise to the dielectric constant \eqref{eq:kappaofG}.

\bibliography{paper}
\bibliographystyle{apsrev4-1}

\end{document}